\documentclass[%
 aip,
 amsmath,amssymb,
 reprint,%
]{revtex4-1}
\usepackage[utf8]{inputenc}

\usepackage{graphicx}
\usepackage{dcolumn}
\usepackage{bm}

\usepackage[utf8]{inputenc}
\usepackage[T1]{fontenc}
\usepackage{mathptmx}
\usepackage{etoolbox}
\usepackage{xcolor}

\draft 

\begin{document}


\title{Towards controlling electron beam charge with nanoparticle-assisted laser wakefield accelerators} 



\author{A. Špádová}
\email{alzbeta.spadova@eli-beams.eu}
\affiliation{ELI Beamlines Facility, The Extreme Light Infrastructure ERIC, Za Radnicí 835, 252 41 Dolní Břežany, Czechia}
\affiliation{Czech Technical University in Prague, Faculty of Nuclear Sciences and Physical Engineering, Břehová 7, 115 19 Prague 1, Czechia}

\author{P. Valenta}
\affiliation{ELI Beamlines Facility, The Extreme Light Infrastructure ERIC, Za Radnicí 835, 252 41 Dolní Břežany, Czechia}

\author{S. Lorenz}
\affiliation{ELI Beamlines Facility, The Extreme Light Infrastructure ERIC, Za Radnicí 835, 252 41 Dolní Břežany, Czechia}

\author{M. Nevrkla}
\affiliation{ELI Beamlines Facility, The Extreme Light Infrastructure ERIC, Za Radnicí 835, 252 41 Dolní Břežany, Czechia}
\affiliation{Czech Technical University in Prague, Faculty of Nuclear Sciences and Physical Engineering, Břehová 7, 115 19 Prague 1, Czechia}

\author{J. Nejdl}
\affiliation{ELI Beamlines Facility, The Extreme Light Infrastructure ERIC, Za Radnicí 835, 252 41 Dolní Břežany, Czechia}
\affiliation{Czech Technical University in Prague, Faculty of Nuclear Sciences and Physical Engineering, Břehová 7, 115 19 Prague 1, Czechia}

\author{G. M. Grittani}
\affiliation{ELI Beamlines Facility, The Extreme Light Infrastructure ERIC, Za Radnicí 835, 252 41 Dolní Břežany, Czechia}

\author{S. V. Bulanov}
\affiliation{ELI Beamlines Facility, The Extreme Light Infrastructure ERIC, Za Radnicí 835, 252 41 Dolní Břežany, Czechia}


\date{\today}

\begin{abstract}
This study explores nanoparticle-assisted electron injection as a method for controlling beam charge in laser wakefield acceleration through particle-in-cell simulations. We systematically investigate how the material (Li through Au) and size (50-200 nm) of nanoparticles influence electron injection dynamics and beam charge. Our results demonstrate that beam charge (10-600 pC) can be effectively controlled by adjusting these parameters. We identify a saturation threshold in the nanoparticle electric field strength, beyond which beam charge depends on the total number of atoms in the nanoparticle rather than on the electron density after ionization. Significant electron injection occurs across multiple plasma wave periods with distribution patterns influenced by nanoparticle properties leading to increased  beam charge but a broader energy spread. 
These findings offer practical guidelines for experimental implementation of nanoparticle-assisted injection in laser wakefield accelerators to tailor electron beam characteristics for various applications.
\end{abstract}


\maketitle 

\section{Introduction}
Laser wakefield acceleration (LWFA) has drawn wide attention from the scientific community as a promising approach for constructing compact high-energy electron accelerators. This mechanism, proposed in 1979 by Tajima and Dawson \cite{Tajima}, uses plasma waves generated via the interaction of intense ultrashort laser pulse with a gas medium to capture and accelerate electrons to relativistic velocities. The main advantage of LWFA is the ability of the plasma medium to sustain very strong electric field (with values up to 100~GV/m) \cite{Esarey}, which enables obtaining electrons with energies up to $\sim$10~GeV with acceleration lengths only several tens of centimeters long \cite{Leemans, Gonsalves_8GeV, Aniculaesei_10GeV, Miao2022, Picksley2024, Rockafellow2025}. Consequently, plasma-based electron accelerators benefit from compact footprints compared to conventional radio-frequency accelerators and have great potential for realizing laboratory-scale accelerators providing high-quality electron beams as sources for strong field quantum electrodynamics \cite{Gonoskov2022, Russell2023, Russell2024}, compact muon sources \cite{Zhang2025, Ludwig2025}, free electron lasers \cite{Wang2021}, radiotherapy \cite{Labate2020, Svendsen2021, Horvath2023, Lazzarini, Guo2025}, or for generation of x-rays covering a broad spectral range \cite{Horny2017, Horny2020, Chaulagain2022, Corde2013, Hojbota2023, Mirzaie2024}.
\par The key to obtaining stable electron beams with high energy and charge, and low energy spread and divergence lies in the electron injection process. Moreover, for reaching multi-GeV electron energies with existing PW and multi-PW lasers, long acceleration lengths and low-density plasmas (10$^{16}$-10$^{17}$ cm$^{-3}$) are required. Under these conditions, most typical injection schemes become difficult to realize, especially in high-power laser guiding configurations and accelerated electron beams can become highly unstable as a result of an unstable injection process. Researchers have proposed several injection mechanisms yielding better electron beam parameters, including the density "down-ramp" injection \cite{Bulanov, Suk2001, Buck2013}, ionization injection \cite{Chen2006, Pak2010, McGuffey2010}, colliding pulse injection \cite{Esarey1997, Faure2006} and their combination \cite{Thaury2015}. Another promising injection schemes use clusters \cite{Fukuda2007}, nanowires \cite{Shen} or nanoparticles \cite{Cho} to trigger the electron injection. Recently, the nanoparticle-assisted laser wakefield acceleration (NA-LWFA), has also been successfully implemented in several experiments \cite{Aniculaesei, Xu, Aniculaesei_10GeV}. In this approach, the nanoparticle contained in the gas target \cite{Lorenz2019, Miao2025} is ionized by the laser pulse, creating a very strong electric field that can attract and subsequently inject electrons from the plasma as well as electrons released by the ionization of the nanoparticle itself.
\par The nanoparticle injection can be achieved even with lower values of the normalized vector potential $a_0$ of the driving laser. Consequently, it enables decoupling of the injection process from the laser evolution and the wakefield generation \cite{Xu}. This is a significant advantage compared to otherwise easily realizable ionization injection, which is heavily dependent on high $a_0$ necessary for the ionization of the inner electron shells. However, with the nanoparticle injection, ionizing lower electron shells is enough and one has the option of choosing materials with lower ionization potential. Most importantly the nanoparticle injection is independent of the rest of the experimental setup, and it can provide stable electron beams with high charge even in low density plasma.
\par Moreover, theoretical studies \cite{Shen, Cho} suggest that by selecting suitable material and size it is possible to control the electron beam parameters. However, electron injection from ionized nanoparticles is a complex process influenced by multiple interdependent factors. Therefore, the same parameters which allow us to tune the beam can also introduce inconsistencies during experiments.
In this work, we investigate the influence of a wide range of nanoparticle parameters on the injected beam charge. We observe clear dependencies on the chosen material as well as the nanoparticle diameter. To systematically explore these effects, we conducted a series of particle-in-cell (PIC) simulations that varied these parameters and analyzed the resulting beam properties. We also discuss the practical aspects of experimental implementation of this injection scheme as it could be a candidate to solve the long standing problem of controlling injection in guided LWFA \cite{Gonsalves_8GeV, Miao2022, Picksley2024, Rockafellow2025}.
\begin{table*}
\caption{Summary of the numerical, laser, and plasma parameters used in the three simulation sets performed with the PIC code Smilei \cite{smilei}. The nppc stands for the number of particles per cell. All simulations use the moving window algorithm.} 
\label{tab1}  
\begin{tabular}{|l|c|c|c|c|c|c|c|c|}
\hline
\rule[-1ex]{0pt}{3.5ex}  simulation set & sim. geometry & grid resolution & grid size & nppc & $a_0$ & $\tau_0$ [fs] & $w_0$ [$\mu$m ]& $n_e$ [cm$^{-3}$]   \\
\hline
\hline
\rule[-1ex]{0pt}{3.5ex}  nanoparticle material & quasi-3D &$\lambda_L/80 \times \lambda_L/20$ & 210 $\times$ 105 $\mu$m$^2$  &16 &  2.8  & 35 &  30 & $5 \times 10^{17}$  \\
\rule[-1ex]{0pt}{3.5ex}  nanoparticle size & quasi-3D & $\lambda_L/40 \times \lambda_L/40$ & 48 $\times$ 20 $\mu$m$^2$ & 16 &2.3 & 6 & 3.5 & $2.1 \times 10^{19}$   \\
\rule[-1ex]{0pt}{3.5ex}  nanoparticle position & full-3D & $\lambda_L/40 \times \lambda_L/20 \times \lambda_L/20$ & 48 $\times$ 48 $\times$ 48 $\mu$m$^3$ & 1 & 2.3 & 6 & 3.5 & $2.1 \times 10^{19}$ \\
\hline
\end{tabular}
\end{table*}

\par The paper is organized as follows. In Sec. II, we describe the setup of the PIC simulations. In Sec. III, we present the results of simulations investigating the nanoparticle material (A) and size (B). In Sec. IV, we discuss the experimental feasibility of NA-LWFA. And finally, in Sec. V, we summarize our findings.

\section{Methods}

\par To study the electron injection and accelerated electron beam parameters, we perform three sets of numerical simulations. The nanoparticle requires at least two cells in every direction to properly resolve its ionization and the created electric field. Therefore, we predominantly use cylindrical (RZ) geometry with azimuthal modes decomposition, as full three-dimensional (3D) simulations are extremely computationally demanding. Since our setup is not far from the cylindrical symmetry (nanoparticle is placed at the simulation axis), for the RZ simulations using only two modes ($m=0$ and $m=1$) was sufficient. We also run a set of 3D simulations where the nanoparticle is located off-axis and the RZ geometry is not sufficient. As is the usual practice for the LWFA simulations we adopt the moving window algorithm in all our simulations. All simulations are done using the PIC code Smilei \cite{smilei}.
\par In the first set of simulations we focus mainly on the nanoparticle material. We simulate a laser with the following parameters: laser wavelength is $\lambda_L = 820$ nm, laser focal spot and pulse duration are $w_0$ = 30 $\mu$m and $\tau$ = 35 fs, respectively (considering Gaussian profile in time and space). The energy in the pulse is 12 J which corresponds to the normalized vector potential $a_0$ = 2.8. The size of the computational domain is 210 × 105 $\mu$m$^2$ and the cell size is $\lambda_L$/80 in the longitudinal direction and $\lambda_L$/20 in the transversal direction. In every cell, eight macroparticles are initialized to model the plasma medium. We consider a plasma with a flat top longitudinal profile with background electron density of $5 \times 10^{17}$ cm$^{-3}$. At the beginning the plasma has a short exponential ramp to avoid artificial plasma wave breaking and laser reflection at the sharp plasma–vacuum interface.

\par The nanoparticle is defined as an initially neutral solid sphere with diameter of 82 nm 
and every nanoparticle cell contains 1~000 macroparticles. The nanoparticle is the main factor determining the resolution. Due to the RZ geometry we need minimum 2 cells per nanoparticle in the transverse direction (which corresponds to $\lambda_L$/20). In longitudinal direction we need to use even finer resolution to resolve the ionization of the nanoparticle. For this purpose we assume only the optical field ionization implemented in Smilei. 
\par Even with the use of the quasi-cylindrical geometry, our simulations are quite computationally expensive; hence, we decide to study only the injection process and the simulations are terminated after approximately 700 $\mu$m of laser propagation. Therefore, we are not able to evaluate the attainable electron beam energy and we focus mainly on the injected charge.
\par The relativistic wavebreaking limit for a plasma wave can be obtained from
$E_{wb} = E_0\sqrt{2(\gamma_w -1)}$, \cite{AP1956, Esarey}
or equivalently from
$E_{wb} =  E_0 a_0 / \sqrt{2}$. \cite{Bulanov1992}
Here $E_0 = mc\omega_p/e$, $m$ is the electron mass, $c$ is the speed of light in vacuum, $\omega_{pe}$ is the electron plasma frequency, $e$ is the electron charge and $\gamma_w$ is the relativistic gamma factor of the plasma wave. The minimum value for the normalized vector potential of the driving laser pulse $a_{wb}$ required to trigger a self-injection is given by $a_{wb} = 2\sqrt{(\gamma_w-1)}$. For conditions used in our simulations $\gamma_{w} \approx \omega_L/\omega_p \approx 57$ and $a_{wb} \approx 15$, where $\omega_L$ is the laser frequency. The value of $a_{wb}$ is clearly much higher than $a_0 = 2.8$, which is used in our simulations. 
Therefore, electrons we observe in the simulations are injected only because of the interaction with the nanoparticle. 

\begin{table}
\caption{Used nanoparticle materials, including the atomic material density $n$ before ionization and the average ionization level obtained from simulations. The specified values of the nanoparticle electron density $n_e$ is after ionization equal to $n$ times the ionization level.} 
\label{tab1}  
\begin{tabular}{|l|c|c|c|c|c|c|c|}
\hline
\rule[-1ex]{0pt}{3.5ex}  nanoparticle material & Li & Al & Ti & Zn & Cu & Ag & Au   \\
\hline
\hline
\rule[-1ex]{0pt}{3.5ex}  ionization level & 3 & 11 & 17 & 20 &  21 & 33 & 45    \\
\rule[-1ex]{0pt}{3.5ex}  at. density $n$ [$\times10^{22}$cm$^{-3}$] & 4.6 & 6.0 & 5.7 & 6.5 &  8.5 & 5.9 & 5.9    \\
\rule[-1ex]{0pt}{3.5ex}  {el. density $n_e$ [$\times10^{23}$cm$^{-3}$]} & 1.39 & 6.63 & 9.62 & 13.1 &  16.1  &  19.3 & 26.5  \\
\hline
\end{tabular}
\end{table}

\begin{figure*}
\includegraphics[scale = 0.92]{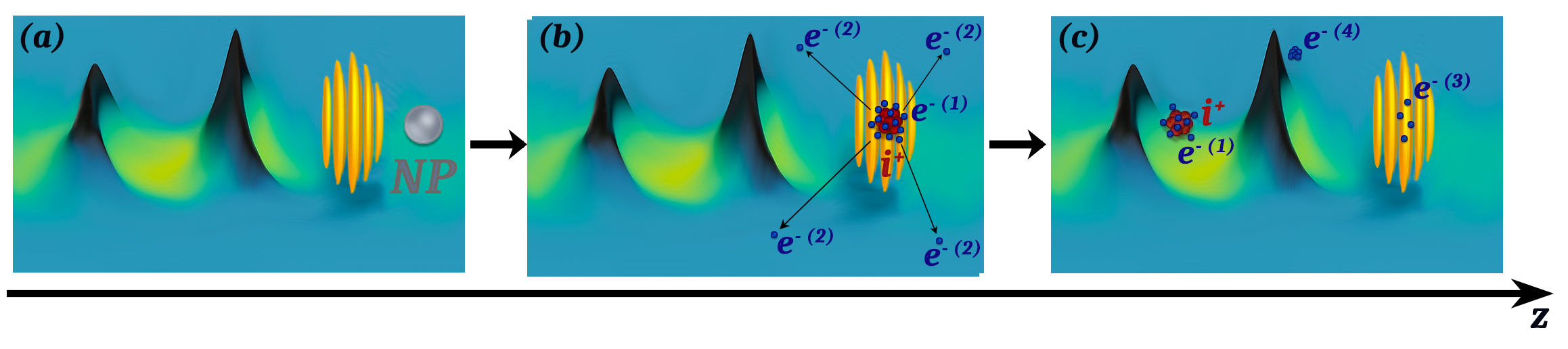}
\caption{\label{fig_scheme}
The scheme of the movement of electrons released from the nanoparticle after its ionization. The laser field is shown with the yellow color, while the electron density (plasma wave) is shown with green-blue-black. Nanoparticle ions and electrons are represented by red and blue dots respectively. (a) The situation before the nanoparticle (NP) ionization. (b) After the nanoparticle interacts with the laser pulse, two groups of electrons are identified: $e^-$$^{(1)}$ are electrons staying close to the nanoparticle ions and $e^-$$^{(2)}$ are electron dispersed into the surrounding plasma. (c) After the first plasma wave period passes the nanoparticle, we recognize two new electron groups: $e^-$$^{(3)}$ which are electrons temporarily trapped inside the laser electric field and $e^-$$^{(4)}$ which are electrons injected in the plasma wave.}
\end{figure*}

\par In the second set od simulations, we explore the impact of the nanoparticle size. Taking into account both, the way the nanoparticle is defined in simulations and the relevant range of the nanoparticle sizes, we need to decrease the size of the computational domain to reduce computational costs. While maintaining as much of the relevant physics as possible. We assume 10 mJ laser with the pulse duration~$\tau =$~6~fs and focus $w_0$ = 3.5~$\mu$m, resulting in the normalized vector potential $a_0$ = 2.3. This choice of laser parameters enables the use of a smaller computational domain of 48 × 20 $\mu$m$^2$ with a relatively small number of cells: 1,920 in the longitudinal direction and 800 in the transverse direction. Here, we had to use a finer resolution in the transverse direction ($\lambda_L/40$) to accommodate for smaller nanoparticles in the simulation. In longitudinal direction, $\lambda_L$/40 was enough to properly resolve the ionization. The electron density in this case is 2.1$\times$10$^{19}$ cm$^{-3}$, which is an optimized value to obtain maximum electron energy for the chosen laser parameters \cite{Valenta}.
\par It should be noted that employing lower laser energies in LWFA implies ultrashort pulse durations, for which the ionization dynamics and subsequent electron acceleration can differ substantially (e.g., due to carrier-envelope phase effects of the laser pulse \cite{Nerush2009, Valenta2020, Huijts2021, Huijts2022}). Therefore, we refrain from making a direct comparison with the previous set of simulations.

\par Finally, to study how the transverse (off-axis) position of the nanoparticle influences the electron injection, we perform additional four simulations in full 3D geometry. This is a situation one will inevitably encounter during experiments, therefore, it is a very important factor to consider. In this simulation set we keep the same laser and plasma parameters as for the size influence study. In the longitudinal direction the resolution is $\lambda_L/40$ and in the transverse directions it is $\lambda_L$/20. The size of the simulation domain is $48 \times 48 \times 48 \mu$m$^3$. The number of macroparticles per cell is decreased from eight to one to model the background plasma while for the 100 nm Al nanoparticle we keep 1 000 macroparticles per cell.

\section{RESULTS}
\subsection{Nanoparticle material}
The chosen materials and their electron densities after ionization are given in Table \ref{tab1}. While the basic injection mechanism is similar across materials, the injection efficiency and beam parameters are significantly influenced by the nanoparticle material. Moreover, as recombination is a process that is orders of magnitude longer than the time period on which the plasma wave interacts with the nanoparticle field we decide to adjust our simulation domain to fit four plasma wave periods. Hence, we are able to observe the nanoparticle field evolution and electron injection in all four plasma wave periods. 

\par We assume the injection mechanism is the same in all four plasma wave periods, i.e., the nanoparticle in plasma induces a localized injection by supplying additional longitudinal momentum to the electrons attracted by the nanoparticle field. Because the nanoparticle field remains strong for longer time than it takes the four plasma wave periods to pass the nanoparticle, there is no reason to expect different injection mechanism. For more detailed description of the injection mechanism, including a mathematical derivation, we refer the reader to paper by Cho et al. \cite{Cho}. 

\subsubsection{Electron injection}
\par First, we observe the evolution of the electrons which are released after the nanoparticle ionization by the driving laser pulse, [Fig. \ref{fig_scheme}(a)]. These electrons are ejected from the nanoparticle with different kinetic energies $E_{kin}$ depending on their ionization potential. Based on the value of $E_{kin}$, they either stay close to the nanoparticle [Fig.~\ref{fig_scheme}(b):~$e^-$$^{(1)}$] or they get dispersed and become a part of the background plasma [Fig.~\ref{fig_scheme}(b):~$e^-$$^{(2)}$]. 
Some of the ejected electrons can also get trapped in the oscillations of the passing laser pulse [Fig.~\ref{fig_scheme}(c):~$e^-$$^{(3)}$]. These electrons are then released later in the simulation. Finally, some of the electrons, which stayed in the vicinity of the nanoparticle, can be injected into the plasma wave and get accelerated [Fig.~\ref{fig_scheme}(c):~$e^-$$^{(4)}$]. 


\par We illustrate this on the example of Cu nanoparticle. The total charge of the ionized Cu nanoparticle with $d$ = 82 nm we obtain from simulations is approximately 85 pC, which is in good agreement with the theory (based on values stated in Table \ref{tab1}). We notice a small fraction around 1.5~pC of the nanoparticle electrons are being trapped in the field of the laser [Fig.~\ref{fig_scheme}(c):~$e^-$$^{(3)}$]. However, if we calculate the negative charge only in the sphere with $r$ = 200 nm with center at the initial position of the nanoparticle we obtain 69.4~pC [Fig.~\ref{fig_scheme}(c):~$e^-$$^{(1)}$]. Therefore, the laser is able to displace [Fig.~\ref{fig_scheme}(b):~$e^-$$^{(2)}$] around 17\% of the total nanoparticle charge. This explains why we observe much lower electric field of the ionized nanoparticle than what would be expected if only a positively charged sphere with the same radius would have been considered.
\begin{figure*}
    \includegraphics[width=0.92\textwidth]{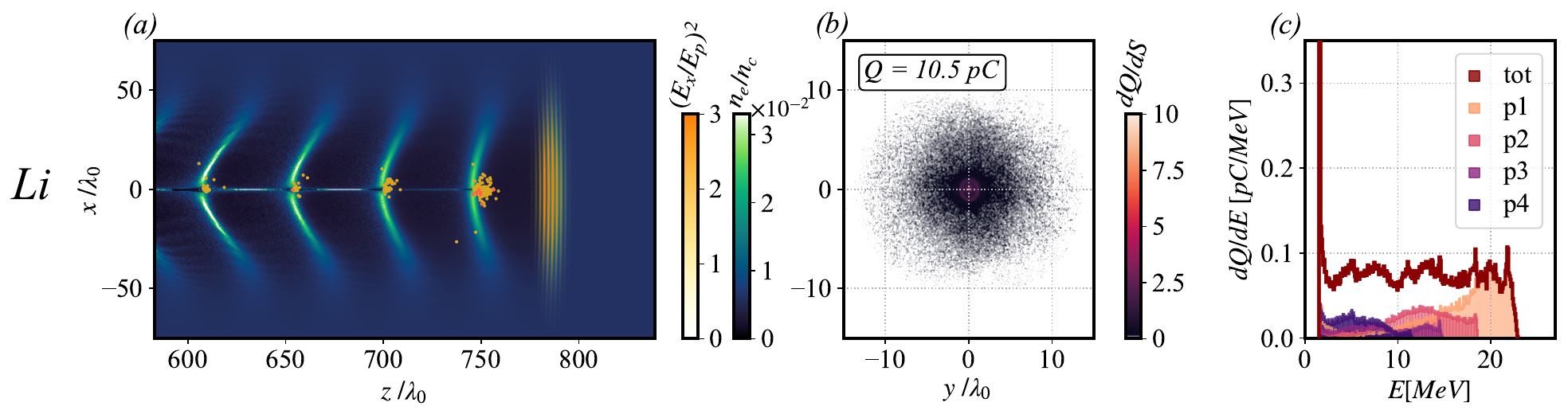}
    \includegraphics[width=0.92\textwidth]{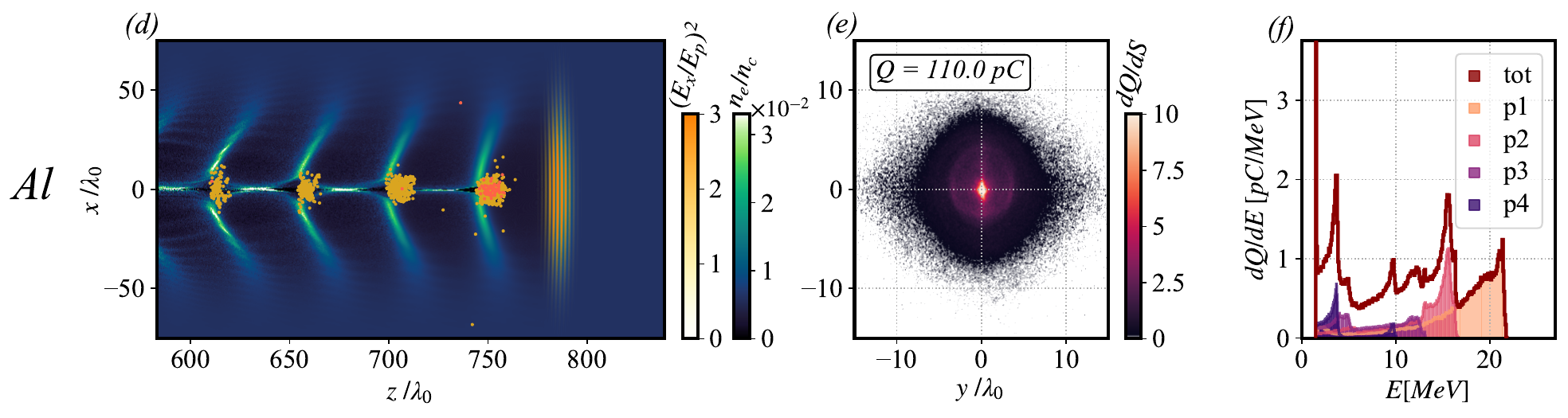}
    \includegraphics[width=0.92\textwidth]{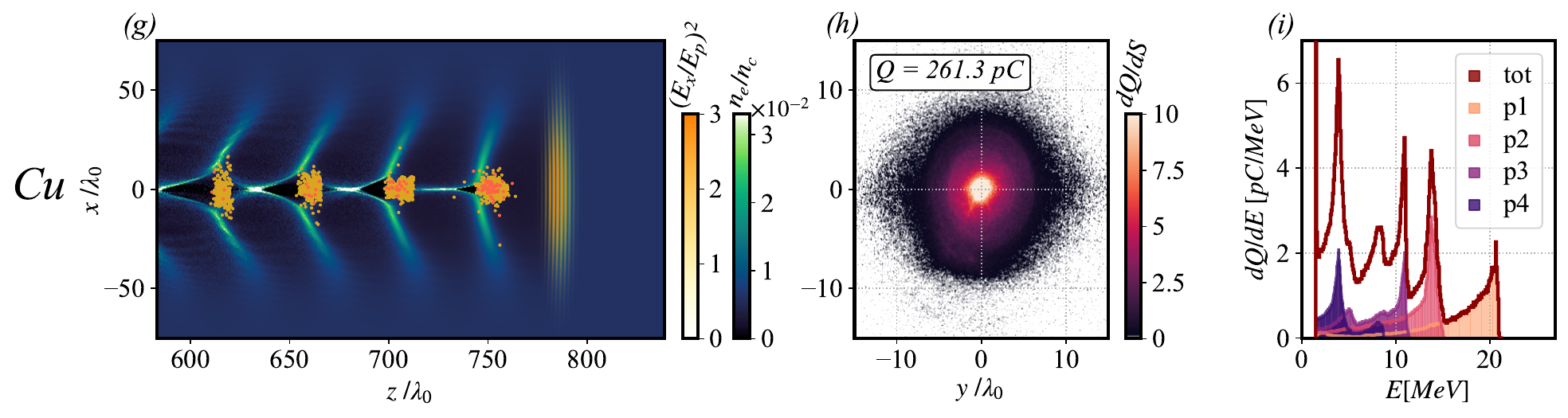}
\caption{\label{fig_wakefield}
Panels (a), (d), and (g) display the laser pulse, the plasma wave and the electron macroparticles depicted by orange (nanoparticle) and yellow (plasma) dots for simulation with Li, Al and Cu nanoparticle, respectively, at $t$ = 710 $T_0$. Panels (b), (e), and (h) show the corresponding electron beam profile together with the injected charge. Panels (c), (f), and (i) depict the corresponding electron spectra, while the spectra obtained from all the four plasma wave periods are represented by the red curve. We also show the spectra from every plasma wave period displayed as the colored area under the red curve. The color scales are saturated.}
\end{figure*}

\par Furthermore, we witness a different behavior based on the nanoparticle material. For, e.g., Al, the charge of approximately 18.7 pC (slightly over 50\% of the total charge after ionization) is also still bound to the nanoparticle. Whereas in case of Li, electrons with the total charge of 7 pC are released after the ionization and only $\sim$14\% of that stays in the vicinity of the nanoparticle ions. Therefore, for Cu and Al nanoparticles, the nanoparticle electric field continues to rise slowly even after the laser pulse passes the nanoparticle. However, in the case of Li the electric field rises faster and the value is not significantly changing during the simulation.
\par We explain this behavior by the influence of the nanoparticle ions which form the nanoparticle field and they apply an attractive force on released electrons. This field is equal to $E = Q / 4\pi \epsilon r^2$, where $Q$ is the charge provided by nanoparticle ions, $\epsilon$ is the vacuum permittivity and $r$ is the distance from the nanoparticle. Hence, the higher the charge of the nanoparticle after ionization, the higher is the attractive force. This force is stronger for the Cu nanoparticle compared to Al or Li. Hence, for higher Z materials ($Z$ is the atomic number) it can be expected that the nanoparticle electric field will be slightly changing within the femtosecond timescale. \\
\par There are two reasons for this behavior, the first is the influence of the plasma wave. As the plasma wave passes the nanoparticle, some nanoparticle electrons can get trapped, or displaced in other way, which causes the growth of the nanoparticle electric field. The second reason is the movement of free electrons nearby the nanoparticle ions. The positive charge of ions is partially shielded by these electrons and as they move, the nanoparticle electric field can vary.

\par Figure \ref{fig_wakefield} shows the end of performed simulations after all the plasma wave periods passed the nanoparticle. Therefore, one can see injected electrons in all of them. We choose three materials (Li, Al, Cu) to compare the injected charge and to determine whether the plasma wave structure is affected by the choice of material. These three materials were chosen to cover a range of various material densities and ionization levels (see Table \ref{tab1}). The electron beam charge clearly follows our expectations; with increasing material density and ionization level the charge is also increasing. 
\par Additionally, one can observe a typical distortion of the plasma wave structure caused when the beam loading \cite{Katsouleas1987} is present. With Li nanoparticle [Fig. \ref{fig_wakefield}(a)], we can see only a hint of beam loading as the injected charge is only 10.5 pC compared to 110 pC for Al [Fig. \ref{fig_wakefield}(d)] and 261 pC in the case of Cu [Fig. \ref{fig_wakefield}(g)]. The overloading limit for a laser with parameters under consideration is approximately 345 pC according to relation $Q^{(b)} = N^{(b)} e$, where $ N^{(b)}\approx \lambda_L / (6\pi r_e) \sqrt{P / \bar P}$. Here the $r_e = e^2 / (4 \pi \varepsilon_0 m_e c^2)$ is the classical electron radius and $\bar P \approx 17.4$ GW is the power characteristic in the theory of relativistic self-focusing \cite{Bulanov2016}. Although we do not reach this limit the injected charge is close enough to this value to create significant modulations in the wakefield. Consequently, we can see the maximum energy at this point of simulation is few MeV higher in the case of Li than in the case of Al and Cu. Hence, we note the beam loading will have quite a significant influence on the maximum attainable energy when using this injection scheme.
\par Panels (c), (f) and (i) of Fig. \ref{fig_wakefield} present the electron spectra. One can see that the maximum reached energy is just over 20~MeV with very high energy spread. This is because the acceleration length at the corresponding simulation time is only about 400~$\mu$m. We expect the total acceleration length with parameters we used would be about few cm or even tens of cm if used in combination with a plasma channel to guide the laser. Note, the dephasing and depletion length in our case is long enough to accommodate for this acceleration distance. Therefore it can be expected that the final electron energy will be somewhere in the GeV range and the energy spread will also improve. Note also the distinctive spectra shape with four peaks - one for every plasma wave period. These peaks are more distinctive for the higher charge beams and for Li nanoparticle they almost diminish. This is caused by the lower injected charge.



\subsubsection{Beam parameters}

\par
\begin{figure}
\includegraphics[width=0.86\linewidth]{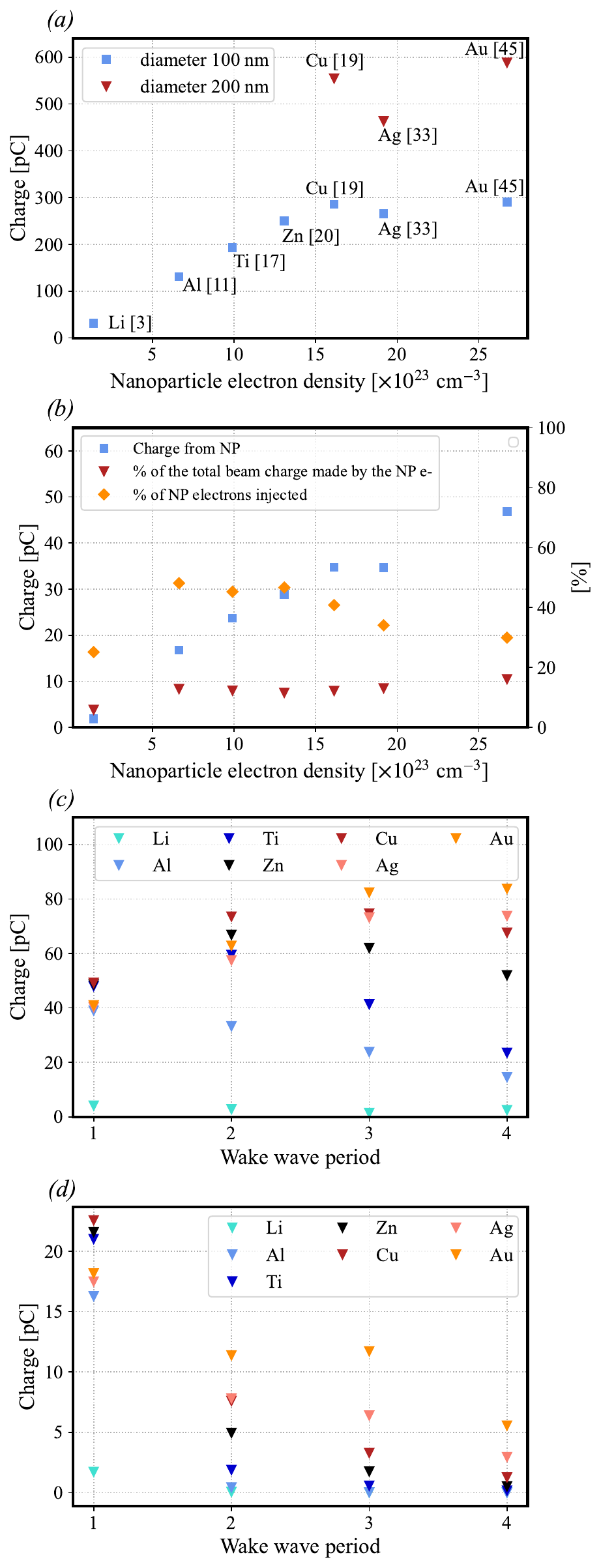}
\caption{ \label{fig:material} 
The effect of the nanoparticle (NP) material on the injected charge of the electron beam. (a)~The beam charge dependency on the electron density of the ionized nanoparticle. In brackets [ ] is the average ionization level of the material atoms. (b) The dependency of the fraction of the beam charge comprising only of the electrons from the ionized nanoparticle on the electron density of the ionized nanoparticle (black) and the percentage of injected electrons created by the nanoparticle ionization (red). (c) The total charge distributed in the first four periods of the plasma wave. (d)~The charge from nanoparticle electrons distributed in the first four periods of the plasma wave.}
\end{figure}

 Next, we investigate the injected charge for all the studied materials, the results are presented in panel (a) of Fig \ref{fig:material}. We are able to obtain qualitatively similar results as in \cite{Cho}. The saturation of the nanoparticle electric field, and hence the injected charge, is notable in our results as well. Cho et al. \cite{Cho} explain this effect by the nanoparticle field saturation. Meaning the strength of the nanoparticle field cannot grow indefinitely and it depends not only on the material but also on the ponderomotive potential of the laser. At first, the material is ionized and electrons are pushed away by the laser. However, once the laser ponderomotive potential equals the electric potential of the nanoparticle, electrons can no longer escape, and the nanoparticle field reaches saturation. However, since we have different laser parameters the saturation density is in our case much higher; around 16$\times 10^{23}$ cm$^{-3}$. 
\par Once the saturation density is reached, we can see that not only the electron density of the ionized nanoparticle is influencing the injected charge but also the density and electron configuration of the material is important. For example the charge injected using Cu nanoparticle is slightly higher than in the case of Ag, even though Ag has larger electron density. While electron density after ionization creates the initial field, the total atom density determines how many ions remain tightly clustered after ionization. Cu with higher atomic density and lower ionization level is able to maintain field strength longer than Ag. This effect is even more distinctive in simulations where the nanoparticle diameter was doubled [red triangles in Fig. \ref{fig:material}(a)]. These simulations are performed to inspect, whether it is possible to inject higher charge with bigger nanoparticles when this is no longer feasible by changing the material. Further details are given in Sec. II B.

\par Moreover, we examine the fraction of the beam charge which is formed by the electrons from the nanoparticle. Similarly to the total beam charge, this quantity is increasing with the nanoparticle electron density [blue squares in Fig. \ref{fig:material}(b)] from few pC up to 50 pC. The percentage of the beam charge which is formed by the nanoparticle electrons is also plotted in the same figure [red triangles in Fig. \ref{fig:material}(b)]. This value stays approximately the same (around 18 \%) for all the studied materials except for Li. Therefore the majority of the beam charge originates from the plasma electrons which gained sufficient momentum for injection from the nanoparticle field. Hence, the beam charge is also significantly dependent on the plasma density. 
\par The last dataset in Fig. \ref{fig:material}(b), yellow diamonds, shows the percentage of electrons created by the ionization of the nanoparticle which are injected into the plasma wave. Omitting the Li point, we can observe a slow decrease form almost 50\% to 37\% with increasing electron density of the nanoparticle. We believe that Li does not follow this trend due to its very low ionization potential. This causes Li to fully ionize even before its interaction with the peak intensity of the laser pulse, allowing its electrons to disperse more widely before the plasma wave reaches the nanoparticle. Consequently, both the Li electrons and the plasma electrons experience a weaker field compared to nanoparticles from higher-Z materials.
\par Furthermore, as the nanoparticle field is diminishing rather slowly we were able to study the distribution of the injected electrons in different plasma wave periods. The results are depicted in panels (c) and (d) of Fig. \ref{fig:material}. Focusing on Fig. \ref{fig:material}(d) one can clearly see that around half of the electrons injected from the ionized nanoparticle is located in the first plasma wave period. Moreover, comparing this value to the Fig. \ref{fig:material}(c), one can see that for Li, Al or Ti these electrons compose the majority of the injected charge. This makes sense as there is a significant amount of electrons present after the nanoparticle ionization and at the same time the weaker electric fields created by nanoparticle with lower electron density attract less electrons from the background plasma. For the rest of the materials the fraction of the charge formed by all electrons/nanoparticle electrons decreases to approximately 25-30\%. If we inspect the material dependence further we can see that for lower electron density nanoparticles, the decline in the number of nanoparticle electrons present in later plasma wave periods is faster than for the ones with high electron density. This phenomenon occurs because the nanoparticle field of high-density materials like Au and Ag diminishes slower, enabling continued efficient injection even in later periods.
\par Even though the charge from nanoparticle electrons is significantly higher in the first plasma wave period compared to the rest, for the most of the materials it is not the highest injected charge in total see Fig.~\ref{fig:material}(c). For most materials the highest charge is injected in the second plasma wave period and as expected for nanoparticles with lower density material the injected charge decreases in latter periods. Note that for Ag and Au the maximum charge is in fact injected in the fourth period. Hence, it is possible that the charge saturation effect is not as strong as it seems. It is likely the injected charge can grow further with the material, however much more slowly. If we include only one plasma wave period in our simulation, we would not have discovered this behavior and we would have obtained almost the same injected charge for all studied materials except for Li.


\subsection{Nanoparticle size}
\par Since the nanoparticle field saturation can be reached quite easily, especially for lower power lasers, we perform another set of simulations focused on nanoparticle size. In these simulations we include six plasma wave periods to confirm the total injected charge is still slowly increasing even after reaching the saturation density. For this purpose we shift from 12 J to 10 mJ laser parameters while maintaining similar normalized vector potential (a$_0$ $\sim$ 2.3). The plasma density is increased appropriately to keep the LWFA regime close to its optimal conditions for achieving maximum electron energies \cite{Valenta}.This approach maintains similar wakefield structures while greatly reducing computational requirements, allowing us to explore a broader parameter space.
Three diameters (50, 100, and 200 nm) and three materials (Li, Al and Cu) were used. 

\begin{figure}
\includegraphics[width=.915\linewidth]{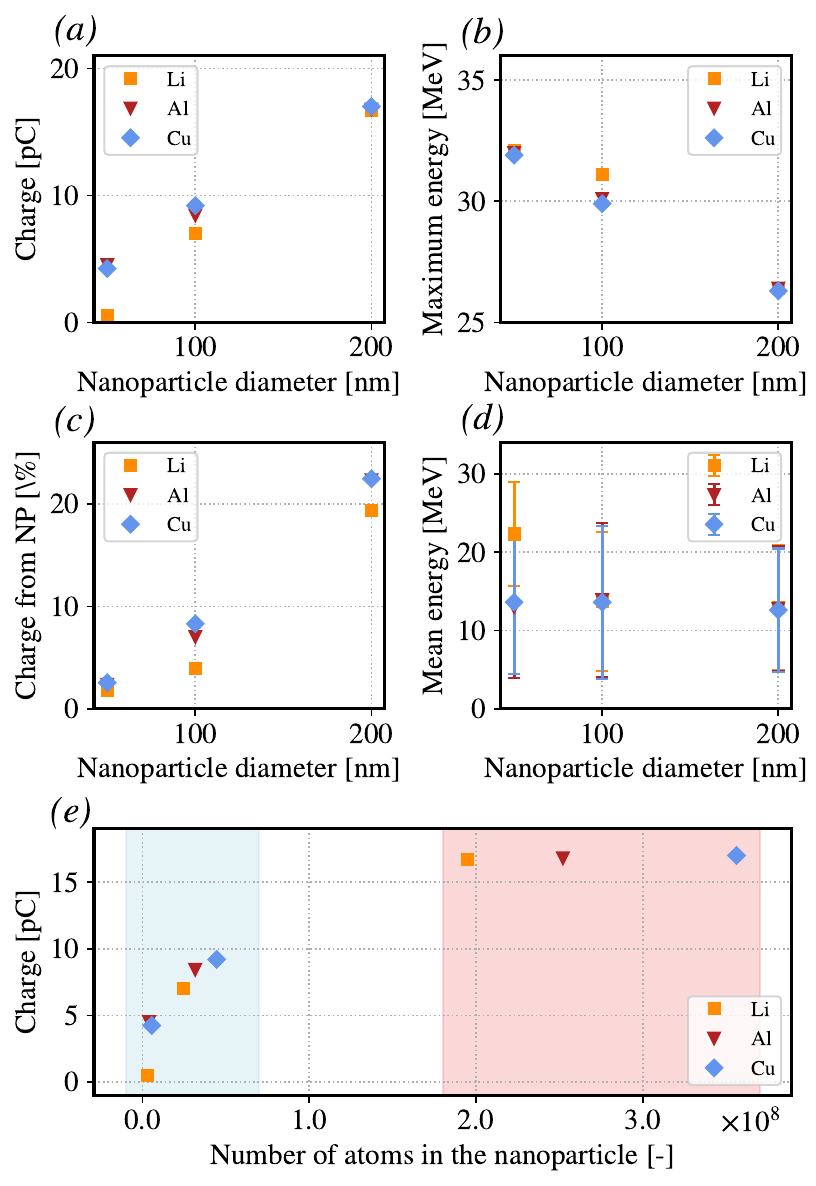}
\caption{\label{fig:size} 
Beam parameters dependency on the nanoparticle parameters (size and material). (a) Charge versus nanoparticle diameter (50-200~nm) for three different materials (Li, Al, Cu). (b) Maximum energy versus nanoparticle diameter. (c) Dependency of the fraction of the beam charge comprising only of the electrons from the ionized nanoparticle. (d) Mean energy dependency on the nanoparticle diameter, with error bars showing the root-mean-square energy spread. (e) Beam charge dependency on the number of atoms forming the nanoparticle (product of its volume and the solid state density). In the blue region solid density of the material still has some influence on the beam charge, in the red region this is no longer important.}
\end{figure} 

\par The simulation results, shown in Figure \ref{fig:size}, reveal clear relationships between the nanoparticle size and electron beam parameters. 
As expected, larger nanoparticle diameters yield consistently higher charge. The dependency in this case seems linear [Fig. \ref{fig:size}(a)] for all used materials. Moreover, there is a notable trade-off between charge and energy [Fig. \ref{fig:size}(b)], with maximum energies around 32 MeV at low charges (i.e., smaller diameters) dropping to about 26 MeV at higher charges (i.e., larger diameters). Clearly there is a significant influence of the beam loading effect which we saw also in the previous set of simulations.
\par The choice of material can also have influence on the energy spread [Fig. \ref{fig:size}(d)]. In the majority of cases, we see the absolute energy spread about 8-10 MeV. The lowest energy spread of 7~MeV we observe for 50 nm Li nanoparticle. Considering the mean electron energy, this yields a relative energy spread of $\sim$30\%, while in the rest of the cases, we see the relative energy spread higher than 65\%. 
This is most probably a consequence of lower injected charge in later plasma wave periods, which is the main phenomenon causing the high energy spread when using this injection scheme. Therefore, in certain cases it may be advantageous to use small nanoparticles and lower density (lower Z) materials to reduce the energy spread.
\par In Fig. \ref{fig:size}(c) we show the portion of the total charge that is formed by the nanoparticle electrons for different diameters. The fraction of the nanoparticle electrons in the accelerated bunch rises from few percents at 50 nm to over 20\% for 200~nm diameter as bigger nanoparticles release more electrons that can be injected.

Finally, in Fig. \ref{fig:size}(e) we show the dependence of the electron beam charge on the number of atoms in the nanoparticle, i.e., we only assume only the nanoparticle size and density at the solid state and we omit its ionization level. Even though we observe a similar ionization levels in this simulation set as with the 12 J laser, the laser is not strong enough to displace more nanoparticle electrons from the ions and create stronger nanoparticle electric field. Therefore the ionization level is no longer relevant but, the solid density still has some influence [blue region in Fig. \ref{fig:size}(e)]. For bigger nanoparticles even the density of the material [red region in Fig. \ref{fig:size}(e)] is no longer important.

\section{DISCUSSION - EXPERIMENTAL FEASIBILITY}

\par In experimental settings, certain constraints limit our ability to freely select nanoparticle parameters. The material selection is the most straightforward to control and offers predictable outcomes but, the range of suitable materials varies with laser power. For PW-class lasers, a wide selection of materials enables effective beam charge tuning. However, for TW-class lasers, finding materials that meet specific requirements becomes challenging, as the saturation density is lower for these lasers. Moreover, as our results indicate, maximizing charge is not always desirable due to beam loading effects.
\par Fortunately, our results demonstrate that nanoparticle size provides an alternative parameter for controlling injected charge. Nanoparticle size can be tuned to some extent e.g. through careful selection of ablation laser parameters, including wavelength, pulse duration, and energy \cite{Kim, Balachandran}. 
The pulse energy can be usually easily tuned even during the experiment. Hence, it is possible to optimize not only the nanoparticle size but also the number of generated nanoparticles, which is a critical parameter to ensure the electron injection in every shot \cite{Xu}. 

\par To fully exploit this injection scheme, it may be necessary to include an aerodynamic lens system (ALS) \cite{Liu1995} in the setup. An ALS is a device able to focus nanoparticles into a tight beam with low divergence. This allows accurate control of the injection location and, consequently, the electron beam energy. Moreover, it provides a precise control over the nanoparticle size, which we are not able to achieve using the laser ablation. The laser ablation technique generates and delivers the nanoparticles into the gas target simultaneously. However, this way one is only able to generate nanoparticles with sizes within a certain distribution \cite{Balachandran} depending on the ablated material and laser parameters. On the other hand, into the ALS enter nanoparticles which were produced beforehand, allowing them to be created with a better control over their size.
\par The biggest challenge for experiments when implementing this injection scheme is controlling the transverse position of nanoparticles during injection. This cannot be solved using an ALS and presents another challenge that cannot be directly addressed during experiments. Therefore, it is crucial to know what to expect if the nanoparticle is not in the ideal position (i.e. not at the center of the wakefield).
\par
\begin{figure}
\includegraphics[width=.95\linewidth]{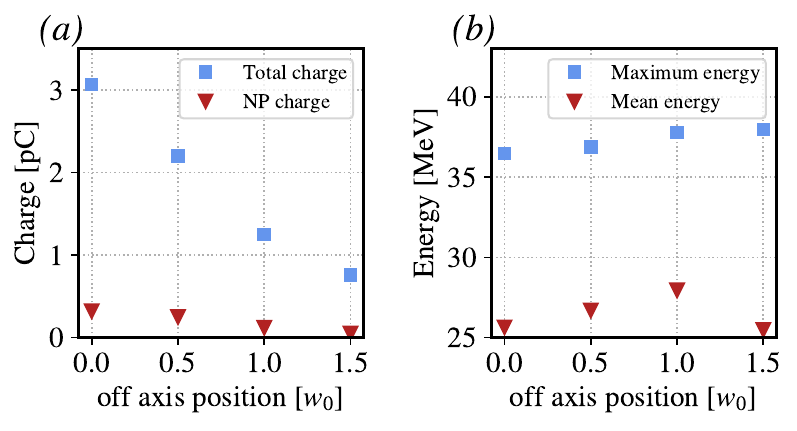}
\caption{\label{fig:displacement} 
Results from the full 3D PIC simulations investigating the influence of transverse nanoparticle position on the beam properties. (a) The total beam charge and the charge formed only by the nanoparticle electrons. (b) Maximum and mean energy for different displacements.}
\end{figure}

 To answer this question we performed additional 4 simulations in full 3D. The results are presented in Figure \ref{fig:displacement}. We observe electron injection even with the transverse displacement larger than the focal spot size although at the cost of decreased beam charge. Note, the reduced charge is not a consequence of the lower laser power nor lower electric field of the nanoparticle as  the strength of $E_{np}$ is almost the same in all simulations. Therefore, we attribute this effect to the weakening of the wakefield with the nanoparticle farther from the simulation axis.

\section{CONCLUSION}

\par In this study, we conducted comprehensive PIC simulations to systematically investigate the influence of nanoparticle properties on electron injection and beam parameters in NA-LWFA. Our findings demonstrate that both the nanoparticle material and the size serve as effective control parameters for tuning electron beam characteristics, particularly beam charge.
\par We confirmed there is a saturation threshold for the nanoparticle's electric field strength that depends on material electron density and the ponderomotive potential of the driving laser. Beyond this threshold, the beam charge dependency shifts from being primarily influenced by the electron density of the ionized nanoparticle to being determined by the total number of atoms in the nanoparticle. This understanding provides crucial insight for optimizing injection schemes in different laser parameter regimes.
\par Importantly, our simulations revealed copious electron injection occurring across multiple plasma wave periods, significantly increasing the total beam charge but also affecting energy spread. The distribution of injected electrons across these periods varies with nanoparticle material and size, offering an additional control mechanism for beam quality.
\par Our results also highlight important experimental considerations, including the trade-off between maximizing charge and maintaining high beam energy due to beam loading effects, and the impact of transverse nanoparticle positioning on injection efficiency. These findings establish clear guidelines for experimental implementation, where careful selection of nanoparticle parameters can be tailored to requirements of a specific application.
\par In conclusion, nanoparticle-assisted injection represents a promising approach for generating stable, high-quality electron beams in laser-plasma accelerators. By providing a detailed understanding of how nanoparticle parameters influence the injection process and resulting beam characteristics, this work contributes to advancing the practical implementation of this technique toward reliable, application-ready accelerators.

\section{ACKNOWLEDGMENTS}

This work was supported by the Ministry of Education, Youth and Sports of the Czech Republic through the e-INFRA CZ (ID:90254).

\par This work was supported by the National Science Foundation and Czech Science Foundation under NSF-GACR collaborative Grant No. 2206059 from the Czech Science Foundation Grant No. 22-42963L.

\section{REFERENCES}

%

\end{document}